\documentstyle[preprint,aps,a4,12pt]{revtex}
\begin{document}
\draft
\hfill\vbox{\baselineskip14pt
            \hbox{\bf KEK-TH-613}
            \hbox{KEK Preprint 99-xxx}
            \hbox{January 1999}}
\baselineskip20pt
\vskip 0.2cm 
\begin{center}
{\Large\bf Quantum Group based Modelling for the description
of high temperature superconductivity and antiferromagnetism}
\end{center} 
\vskip 0.2cm 
\begin{center}
\large Sher~Alam\footnote{Permanent address: Department of Physics, University
of Peshawar, Peshawar, NWFP, Pakistan.}
\end{center}
\begin{center}
{\it Theory Group, KEK, Tsukuba, Ibaraki 305, Japan }
\end{center}
\vskip 0.2cm 
\begin{center} 
\large Abstract
\end{center}
\begin{center}
\begin{minipage}{14cm}
\baselineskip=18pt
\noindent
 Following our recent conjecture to model the phenomenona of
 antiferromagnetism and superconductivity by quantum symmetry
 groups, we propose in the present note three toy models, namely,
 one based on ${\rm SO_{q}(3)}$ the other two constructed with
 the ${\rm SO_{q}(4)}$ and ${\rm SO_{q}(5)}$ quantum groups. 
 Possible motivations and rationale for these choices are 
 outlined. One of the prime motivations underlying our proposal
 is the experimental observation of {\em stripe} structure [phase]
 in high T$_{\rm c}$ superconductivity [HTSC] materials. A number 
 of experimental techniques have recently observed that the 
 ${\rm CuO_{2}}$ are rather inhomogeneous, providing evidence for 
 phase separation into a two component system. i.e. carrier-rich 
 and carrier-poor regions. 
 In particular, extended x-ray absorption fine structure [EXAFS] 
 demonstrated that these domains forms stripes of undistorted and 
 distorted local structures alternating with mesoscopic length scale
 comparable with coherence length in HTSC.
\end{minipage}
\end{center}
\vfill
\baselineskip=20pt
\normalsize
\newpage
\setcounter{page}{2}
	
	The Hubbard Hamiltonian [HH] and its extensions dominate
the study of strongly correlated electrons systems and the 
insulator metal transition \cite{fra91}. One of the attractive
feature of the Hubbard Model is its simplicity. It is well known
that in the HH the band electrons interact via a two-body
repulsive Coulomb interaction; there are no phonons in this model
and neither in general are attractive interactions incorporated.
With these points in mind it is not surprising that the HH
was mainly used to study magnetism. In contrast superconductivity
was understood mainly in light of the BCS theory, namely as
an instability of the vacuum [ground-state] arising from
effectively attractive interactions between electron and 
phonons. However Anderson \cite{and87} suggested that
the superconductivity in high T$_{c}$ material could arise
from purely repulsive interaction. The rationale of
this suggestion is grounded in the observation that
superconductivity in such materials arises from the
doping of an otherwise insulating state. Thus following this
suggestion the electronic properties in such a 
high T$_{c}$ superconductor material close to a 
insulator-metal transition must be considered.
In particular the one-dimensional HH is considered
to be the most simple model which can account
for the main properties of strongly correlated
electron systems including the metal-insulator 
transition. Long range antiferromagnetic order
at half-filling has been reported in the numerical
studies of this model \cite{hir89,bal90}.
Away from half-filling this model has been studied
in \cite{mor90-91}.

	In theories based on magnetic interactions \cite{and87}
for modelling of HTSC, it has been assumed that the CuO$_{2}$
planes in HTSC materials are microscopically homogeneous.
However, a number of experimental techniques have recently observed 
that the CuO$_{2}$ are rather inhomogeneous, providing evidence for 
phase separation into a two component system. i.e. carrier-rich and 
carrier-poor regions \cite{oyn99}. In particular, extended x-ray 
absorption fine structure [EXAFS] demonstrated that these domains 
forms stripes of undistorted and distorted local structures alternating 
with mesoscopic length scale comparable with coherence length in HTSC.
The neutron pair distribution function of Egami et al. \cite{ega94}
also provides structural evidence for two component charge carriers.
Other techniques also seem to point that below a certain temperature
T$^{*}$ \footnote{The following can be taken as a definition of T$^{*}$:
T$^{*}$ is an onset temperature of pseudogap opening in spin or
charge excitation spectra.} the CuO$_{2}$ planes may have ordered stripes
of carrier-rich and carrier-poor domains \cite{ega94}. The emergence
of experimental evidence for inhomogeneous structure has led to
renewal of interest, in theories of HTSC which are based on 
alternative mechanism, such as phonon scattering, the lattice
effect on high T$_{\rm c}$  
superconductivity \cite{lat92,phs93,phs94,ega94}.
Polarized EXAFS study of optimally doped YBa$_{2}$Cu$_{3}$O$_{\rm y}$
shows in-plane lattice anomaly \cite{oyn99} below a characteristic 
temperature T$^{*'}$\footnote{T$^{*'}$ may be defined as follows: 
T$^{*'}$ is an onset temperature of local phonon anomalies
and T$^{*'} < T^{*}$.} which lies above T$_{\rm c}$, and
close to the characteristic temperature of spin gap opening
T$^{*}$. It is an interesting question if the in-plane
lattice anomaly is related to the charge stripe or spin-phonon
interaction. We note that it has been attempted in \cite{fuk92,fuk93}
to relate the spin gap observed in 
various experiments such as NMR, neutron scattering and transport
properties to the short-range ordering of spin singlets.

	Zhang \cite{zha97} proposed a unified theory
of superconductivity and antiferromagnetism 
\footnote{We note that theories of cuprates based 
upon a quantum critical point have also been suggested
by others, see for e.g. Sachdev et al. \cite{sac95}.}based
on SO(5) symmetry and suggested that there exists
an approximate global SO(5) symmetry in the low
temperature phase of the high $T_c$ cuprates.
In this model one has a five component order
parameter. Three components correspond to a spin
one, charge zero particle-hole pair condensed at the
center of mass momentum $(\pi,\pi)$, these components
represent antiferromagnetic order in the middle of 
Mott insulating state. The remaining two components
correspond  to a spin-singlet charge $\pm 2e$
Cooper pair of orbital symmetry $d_{x^2-y^2}$
condensed in zero momentum state, these last
two components are supposed to correspond to
superconductivity in the doped Mott insulator.
Baskaran and Anderson \cite{bas97} have presented
an elegant series of criticisms of the work in Ref.~\cite{zha97}. 
In view of the critique presented in Ref.~\cite{bas97},
and the features of quantum groups we were led to our 
conjecture that at the simplest level and as a
preliminary step, it is tempting to base a model
for superconductivity and antiferromagnetism on
a quantum group symmetry rather than the usual
classical Lie group \cite{ala98}.

 	We now recall some useful details about quantum
groups \cite{kli97,kak91} and our simple discussion about
quantum groups outlined in \cite{ala98}. We caution the 
reader that currently there is no `satisfactory' general 
definition of a quantum group\footnote{We mean in terms of rigorous
mathematics}. However it is commonly accepted \cite{kli97}
that quantum groups are certain `well-behaved' Hopf
algebras and that the standard deformations of the enveloping
Hopf algebras of semisimple Lie algebras and of coordinate Hopf
algebras of the corresponding Lie groups are guiding examples.
An amazing feature of quantum group theory is the 
unexpected connections with apparently unrelated
concepts in physics and mathematics such as Lie Groups,
Lie Algebras and their representations, special functions,
knot theory, low-dimensional topology, operator algebras,
noncommutative geometry, combinatorics, quantum inverse
scattering problem, theory of integrable models, conformal
and quantum field theory and perhaps others.

	 We consider even the modelling of HTSC 
materials based on quantum groups as a preliminary step, since 
it is quite likely that a realistic model which unifies a complex
system containing antiferromagnetic and superconducting
phases, may require the mathematical machinery
currently being used for string theory, or something even 
beyond it. Another important motivation for our
conjecture is to model Stripes. Stripes can be aptly described
as being found in the unstable two-phase region between the 
antiferromagnetic and metallic states \cite{bas97-1}. One
of the real challenges is to formulate a theory which gives rise 
to the equivalent of ``Fermi surface''\footnote{We mean, as is 
usual \cite{bas97-1}, that with a metallic state we can always
associate a ``Fermi surface'' which describes the low-energy
excitations. This is a finite volume surface in momentum space 
which arises due to all the one-particle amplitudes \cite{bas97-1}.} 
whose excitation surface derives from {\em fluctuations that 
are not uniform in space}. Intuitively one may imagine
these non-uniform fluctuations as arising out of a
nonlinear sigma-like model. We further conjecture that
superconductivity arises when two immiscible phases,
namely a 2-D antiferromagnetic state and a 3-D metallic state,
are ``forced'' to meet at $T_c \rightarrow \infty$.
As is well-known, ordinary low temperature superconductors 
arise entirely out of a ``metallic'' like  state.
In contrast High $T_{c}$ superconductors have
relatively large $T_{c}$ since we {\em cannot}
smoothly map two immiscible states [metallic and
antiferomagnetic] together. It is the lack of this
smooth mapping that is precisely responsible for
the High $T_c$. The lack of smooth mappings may
be modelled by the nontrivial phases which
arise out of the braiding operations of
quantum groups. 

	In lieu of our remarks on quantum groups \cite{ala98}
to model antiferromagnetism and superconductivity
we first consider a simple toy model, namely SO$_{\rm q}(3)$.
Why SO$_{\rm q}(3)$?  To this end we recall the observation that
superconductivity in HTSC materials arises from the
doping of an otherwise insulating state. On the other hand
it is known that the low energy effective Hamiltonian of
a antiferromagnet can be described by the SO(3) nonlinear
$\sigma$ model \cite{cha89}. There are also claims 
\cite{pin94,pin95,chu94} that the SO(3) nonlinear sigma model may
also suffice to describe the underdoped region of
the oxide superconductors. Simply put, experimentally we know
that the doping of oxide antiferromagnet leads to superconductivity 
for a region of the doping parameter. The actual picture is more
complicated for example by the presence of stripes. To be
concrete we have in mind the generalized phase diagram
as seen in  La-SrCuO$_{4}$ [see Fig.~3.26, page 98 of 
Ref.~\cite{and97}, where the following phases are
enumerated: antiferromagnetic insulator, messy insulating
phase, 2D strange metal, 3D metal and superconductor on
the temperature versus doping diagram.].
In a simple toy model we may model the antiferromagnetic
material plus the doping by ${\rm SO_{q}(3)}$, where the
quantum group parameter $q$ is related to the physical doping
$\delta$. In a naive scenario we may take $\delta=q-1$, when $q=1$
${\rm SO_{q}(3)}$ reduces to ${\rm SO(3)}$, $\delta=0$ 
i.e. zero doping and we recover the SO(3) nonlinear
$\sigma$ model description for the low energy effective
Hamiltonian.   

	There are several reasons for choosing  
${\rm SO_{q}(4)}$ and ${\rm SO_{q}(5)}$
As already mentioned SO(3) is a symmetry group
for antiferromagnet insulator at the level of effective
Hamiltonian. On the other hand effective Hamiltonian
of a superconductor may be described by a U(1) nonlinear
sigma model [XY model], for example it was indicated
in Ref.~\cite{don90} that the metal insulator transition may
be described by the XY model. It was further pointed out
in  Ref.~\cite{eme95} that superconducting transition
on the underdoped side of oxides by a renormalized
classical model. In addition it is worth noting that
SO(3) spin rotation and U(1) phase/charge rotation
are symmetries of the microscopic t-J model.
One of the simplest group to embed SO(3)$\times$ U(1)
is ${\rm SO(5)}$. Now keeping in mind the discussion
in \cite{ala98} and here we propose to model 
a unified group for antiferromagnetic, superconducting
and other phases in HTSC materials by ${\rm SO_{q}(5)}$.
Even in the considerations based on SO(5) the concepts of Hopf
maps, quaternions, Yang monopole and Berry phase \cite{zha98} 
have crept up. This leads more support to our contention
that model for HTSC materials must be based on quantum group
rather than the classical group in this case ${\rm SO_{q}(5)}$
instead of SO(5). A singlet-triplet model has been suggested
recently by Mosvkin and Ovchinnikov \cite{mos98} in light
of experimental considerations. This model is based on
the Hamiltonian of the two-component spin liquid \cite{mos98}.
The Hamiltonian considered in \cite{mos98} has SO(4)
group symmetry. The calculations and results in \cite{mos98}
are encouraging in that they elaborate and give insight
into static and dynamical spin properties of cuprates
including paramagnetic susceptibility, nuclear resonance
and inelastic magnetic neutron magnetic scattering.
However a central issue as recognized in \cite{mos98}
and not dealt with is charge inhomogeneity and
phase separation. It is tempting to replace SO(4)
by ${\rm SO_{q}(4)}$ and see if one could account
for charge inhomogeneity and phase separation.

	We must also clearly state that we don't at the moment
have a magic insight to tell us which specific quantum group
must be chosen to model HTSC. However we feel that the quantum groups
arising from the classical orthogonal groups, i.e. SO(N)
are a good and worthwhile starting point, since they
naturally incorporate the symmetry group of the insulating 
antiferromagnetic state and are naturally rich enough
to accomodate quantum liquid behaviour. 

    In conclusion, the quantum orthogonal groups 
${\rm SO_{q}(N)}$ are proposed as potential candidate 
for modelling the theory of HTSC materials. A strong
feature of quantum groups is that they unify classical
Lie algebras and topology. In more general sense it
is expected that quantum groups will lead to a deeper
understanding of the concept of symmetry in physics. 
 
\section*{Acknowledgments}
The author's work is supported by the Japan Society for
the Promotion of Science [JSPS]. 

\end{document}